\documentstyle[10pt,epsf,fleqn]{elsart}
\newcommand{\On}{0$\nu\beta\beta$-decay}
\newcommand{\Tn}{$\beta\beta$-decay}
\newcommand{\GE}{$^{76}$Ge}

\def \be {\begin{equation}}
\def \ee {\end{equation}}
\def \KK{H.V.~Klapdor-Kleingrothaus}
\def \HdMo{Heidelberg-Moscow-Experiment}

\begin{document}
\begin{frontmatter}
\title{Digital Pulseshape Ana\-lysis by Neural Networks for the
  Heidelberg-Moscow-Double-Beta-Decay-Experiment}
\author{B.~Majorovits, \KK}
\address{Max-Planck-Institut f\"ur Kernphysik, Postfach 103980, 69029 
 Heidelberg, Germany}

\begin{abstract}
The Heidelberg-Moscow Experiment is presently the most sensitive
experiment looking for neutrinoless double-beta decay. Recently the
already very low background has been lowered by means of a Digital
Pulseshape Analysis using a one parameter cut to distinguish
between pointlike events and multiple scattered events. To use all the
information contained in a recorded digital pulse, we developed a new
technique for event recognition based on neural networks.
\end{abstract}
\end{frontmatter}

\section{Introduction}
The question of a nonvanishing neutrino mass is still one of the most
outstanding open problems in modern physics. Especially after the
latest striking hints for neutrino oscillations from the
Super-Kamiokande experiment \cite{SuperK} it has become very important
to verify these results independently.  
Neutrinoless double-beta (0$\nu\beta\beta$) decay, which
violates
Lepton-number and B-L conservation by two units,
 is one of 
the most promising tools for the search of a finite neutrino mass and
some other physics beyond the standard model \cite{KLAP}. Furthermore 
it seems to be the only possibility to distinguish between the
Majorana- and the Dirac-nature of the neutrino.
If \On~is observed neutrinos have to be of
Majorana-type and have a finite mass. The
atmospheric neutrino problem confirmed by the Super-Kamiokande
collaboration \cite{SuperK} has brought degenerate neutrino models
back to 
attention again \cite{valle}, where all neutrinos have a mass in the
order of $\mathcal{O}$(eV). 
The newest generation of \On~
experiments, especially the \HdMo \cite{HdMo} started already now to test this 
mass range.

\section{The Heidelberg-Moscow-Experiment}
The \HdMo~ is presently the most sensitive experiment looking for
\On~\cite{HdMo}. Out of 19.2~kg enriched \GE~five p-type High-Purity
Germanium (HPGe) crystals
were grown, which are now operated as p-type detectors in the Gran
Sasso Underground Laboratory with an active mass of 10.96~kg in an
extremely radiopure surroundings.  
The experiment has a background rate of 0.2~counts/(kg~keV~y) 
in the energy region between 2000~keV and 2080~keV,
where the expected signal, a sharp peak at 2038.56~keV (the Q-value of
\Tn) lies. 
Since 1995 an additional background reduction has been achieved through
the use of Digital Pulseshape Analysis (PSA). Due to the fact that the
shape of the detected pulse is dependent on the type of interaction a
distinction between multiple scattered
Compton events and single interaction events (as \On) is possible.
A \On~event would appear as a Single Site Event (SSE), since
the mean free path of the two electrons emitted by the decay is
smaller than the time resolution of the detector allows to 
distinguish due to the low drift
velocities of the electron-hole pairs. This means that Multiple Site
Events (MSE) in the energy region of \On~can be regarded as background. 
To distinguish between the two interaction types a one-parameter
method was developed at that time, based on the fact that the time 
structure of the pulse shapes in
Germanium detectors are mainly dependent on the locations of
the various events of a count within the HPGe-crystal. For MSE's one
therefore expects a broader pulse in time than for SSE's
since the initial locations of the electron-hole pairs are distributed
over a larger area of the crystal and the overall detection time
therefore increases.
With this method a reduction of the background  by a factor of three
in the area of the expected signal could be reached \cite{Jochen,PhysLettB}.

Nevertheless a large amount of information is neglected with this method 
since only one parameter serves as the distinguishing criterium. 
Furthermore the method relies on a statistical correction of the
measured SSE pulses 
since the efficiency of the method is substantially smaller than 100\%
resulting in a loss of information about the single events.
For this reason we developed a new method based on neural networks to
use as much information as possible from the
recorded pulse shapes and to avoid statistical treatment of the
obtained data.

\section{Neural Networks}
Neural Networks are nowadays used in a wide variety of applications like 
pattern-, image- and videoimage-recognition.
Since in the case of PSA the discrimination technique relies on a sort of
pattern recognition it seemed consequent to base a new PSA-technique
on this method. In contrast to the old method, where only one
parameter was used as the distinguishing criterium, all the information 
obtained by the measurement about the time structure of the pulse
is fed to the neural
networks in order to distinguish between SSE's and MSE's.

\begin{figure}[t]
\hspace*{3cm}
\epsfxsize8cm
\epsffile{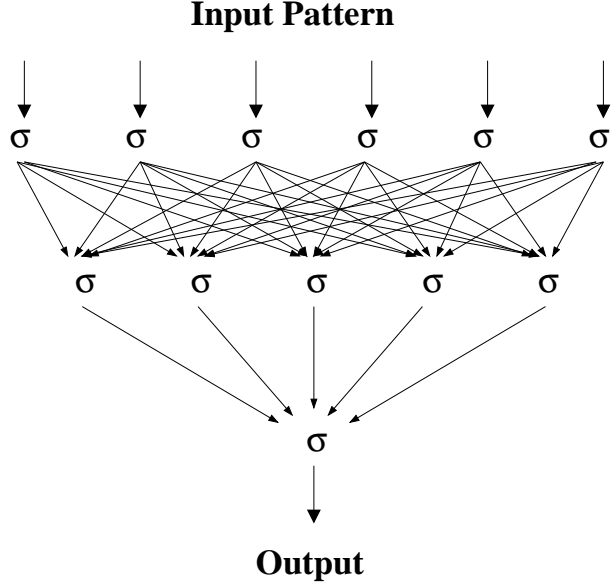}
\caption{\label{perceptron} Design of a feed forward Neural Network with three
  layers: one input layer, one hidden layer and one output layer. The
  $\sigma$ at each knot represents the neuron's activation function
  which is of sigmoid type in our example.}
\end{figure}

Typically a network is divided into processing units, which are
further divided into single neurons. 
Each unit recieves signals from the previous level (i.e. from the neurons in 
the unit) and computes an output, which is
then passed further to the next unit (i.e. to the neurons of the
unit). The schematic action of such a feed forward
network is depicted in Fig. \ref{perceptron}.

A typical neural network consists of three layers: the input layer,
the hidden layer and the output layer.
It has been shown that such a network suffices to approximate any
function with a finite number of discontinuities to arbitrary precision if the
activation function of the hidden unit neurons are nonlinear
\cite{NetApprox}.

If one has digitized information in an array (in our case it is the time
evolution of the measured current behind the preamplifier,
i.e. it is one-dimensional), the entries $x_{j}$ can be passed to the
input layer to 'activate' the neurons through the activation function,
typically of the sigmoid form
\begin{equation}
\begin{centering}
{\cal F}(x_{j})=y_{i}(x_{j})=\frac{1}{1+e^{-x_{j}}}.
\end{centering}
\end{equation}
Each neuron then passes its activation value $y_{i}$ to all
the neurons in the hidden layer after multiplying it with a weight
factor, so that the input to the neurons in the next layer is given by:
\begin{equation}
\begin{centering}
x^{h}_{j}=\sum_{i} w_{ij}y_{i}+\theta_{j}
\end{centering}
\end{equation} 
where $\theta_{j}$ is a threshold specific to the layer and $w_{ij}$ is
the corresponding weight between the i-th neuron in the input layer
and j-th neuron in the hidden layer.
Again the output of the neuron is calculated through the activation
function given in (1) and passed to the neurons of the next layer.
Finally one obtains the output signals from the activation of
the output neurons.
Often (like in this analysis) the output layer consists of only one
neuron returning a value  
between 0 and 1 thus deciding whether the data passed to the input
layer belongs to a signal of type A) or B).

However the network has to be configured in order to be able to
distinguish reasonably between two types of input patterns. This is
mostly done by a sort of training process. If one has a library of input 
patterns, these can be passed to the network. After the input pattern
has been applied and the output has been calculated, the connections
between the neurons are adjusted according to the generalized delta rule:
\begin{equation}
\Delta w_{ij}=\gamma\delta_{j}y_{i},
\end{equation}
where $\gamma$ is the learning rate, $y_{j}$ is the activation of the
neuron due to the given input pattern and $\delta_{k}$ is an error
signal which in our case (sigmoid activation function) is given for
the output layer by
\begin{equation}
\delta_{o}=(d_{o}-y_{o}){\cal F}'(x_{o})=(d_{o}-y_{o})y_{o}(1-y_{o})
\end{equation}
and by
\begin{equation}
\delta_{h}={\cal F}'(x_{h})\sum_{o=1}^{N_{0}}\delta_{o}w_{ho}=y_{h}(1-y_{h})\sum_{o=1}^{N_{0}}\delta_{o}w_{ho}
\end{equation}
for the hidden layer. Here ${\cal F}'$ corresponds to the first
derivative of the activation function, $d_{0}$ is the expected
result of the output neurons and N$_{o}$ is the number of output neurons.
Often, like in this analysis, a momentum term is used in the learning
process to avoid oscillations in the training procedure:
\begin{equation}
\Delta w_{jk}(t+1)=\gamma\delta_{k}y_{j}+\alpha\Delta w_{jk}(t),
\end{equation}
where $t$ is the presentation number and $\alpha$ is a constant
representing the effect of the momentum term.

After a certain number of these training procedures the network
'learns' the patterns of the types of input information and the output 
of the network results in a value close to zero for a pattern
of type A) and in a value close to one for a pattern of type B).

For a general introduction to Neural Networks see for example \cite{Networks}.

\begin{figure}[h]
\hspace*{2.5cm}
\epsfxsize10cm
\epsffile{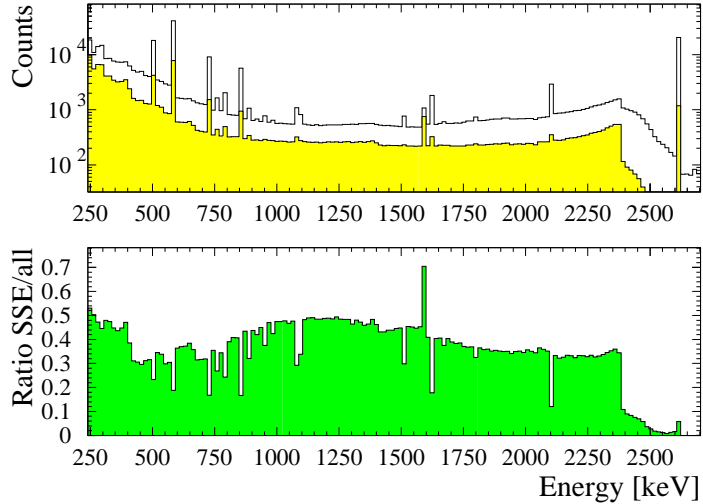}
\caption{\label{spectrum_ratio} Expected result of PSA from the
  simulation of the $^{228}$Th- calibration spectrum. Upper panel:
  Simulated spectra of all events (open 
  histogram) and SSE events only (shaded histogram). Lower panel:
  Simulated ratio of SSE's in the spectrum as a function of energy. In 
  the energy region of \On~a reduction by a factor of $\sim$2.8 can be
  expected.}
\end{figure}

\section{Digital Pulsform Analysis with Neural Networks}
In order to perform PSA a sufficiently large library of known
reference pulses has to be collected for the training process.
A reliable source of the two different kinds of pulses is needed for
this reason.
It is well known that high energetic (E $>$ 500 keV) total absorption
peaks consist mainly of Compton scattered events
(see. \cite{Roth}). The amount of MSE's in these peaks in general 
is not less than 80\%. In contrast to this, Double-Escape peaks (pair
production followed by the annihilation of the e$^{+}$ and the escape
of both 511 keV $\gamma$'s) consist of SSE's only, since the detected
particle in this case is a single electron with energy
\begin{equation}
E_{DE}=(E_{0} - 2\times511~keV), 
\end{equation}
whose dissipation length again
is smaller than the time resolution of the detector allows to
resolve. Only the Compton
background from higher energetic peaks in this area contributes to a
conta\-mination of MSE's in the peak region of the Double-Escape line.
Using a $^{228}$Th-calibration source, the
Double-Escape line of the total absorption peak at 2614.53~keV with an
energy of E$_{DE}$=1592.5~keV can be used for the SSE sample.
To avoid systematic effects in the training process, a total
absorption peak with a similar energy should be used for the MSE
sample. The peak at 1621~keV from the $^{228}$Th-daughter nuclide
$^{212}$Bi seemed appropriate for this purpose.

\begin{figure}[h]
\epsfxsize10cm
\hspace*{2.5cm}
\epsffile{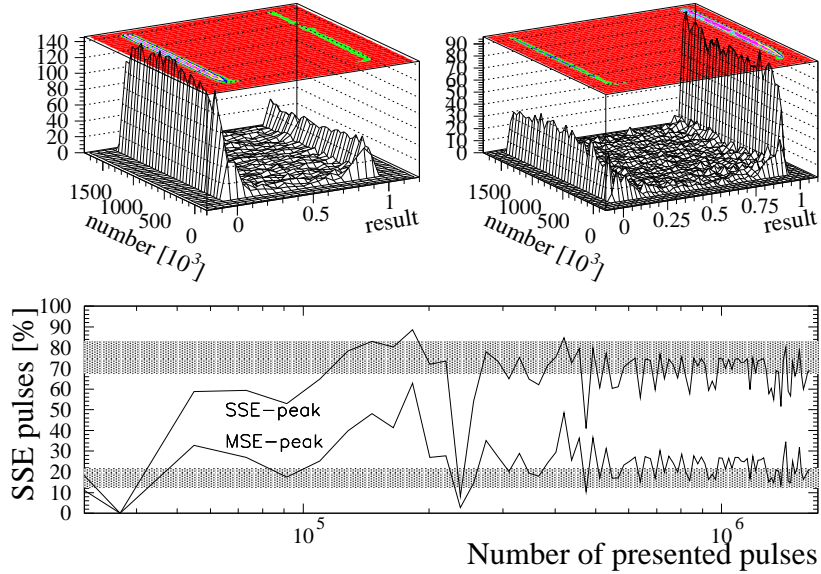}
\caption{\label{net_evolution} Evolution of the network during the
  training process. The result of the network after feeding it 
  training pulses was monitored during the training process. Upper
  panel: The  Events with the results from the network are shown as a
  function 
  of the number of presented pulses. Left: Result from SSE-sample
  (1592~keV) Right: Same for MSE-sample (1621~keV).
  Lower panel: Fraction of identified SSE's in a given sample of
  reference pulses as a function of number of presented pulses. The
  shaded areas correspond to the expectation (one sigma band) obtained
  from the simulation.
  }

\end{figure}

\section{Simulation of PSA}
To test the efficiency and the reliability of the new method we
performed simulations of calibration measurements of the \HdMo. 
It is especially important to check for a possible energy dependence
of the method since the energy of the training pulses ($\sim$ 1.6 MeV) 
does not
coincide with the energy region of the expected \On~signature
(2038.5~keV). For this purpose we used the 
GEANT3.21 Monte-Carlo code \cite{GEANT} extended for low energetic decays. The
geometry of the 
experiment and the library of low energetic decays was programmed and
successfully tested in earlier works \cite{Maier,Dietz}. The code was 
further extended to distinguish between multiple and single
interaction events \cite{Gund}.
In Fig. \ref{spectrum_ratio} the simulated spectra of a calibration
measurement with a $^{228}$Th source with the whole setup of the
\HdMo~and the resulting 
expected ratio of SSE's in the spectrum are shown. 
In the energy region of the \On~ between 2000~keV and 2080~keV a
reduction factor of $\sim$2.8 is expected through the use of PSA. 

\begin{figure}[h]
\hspace*{3.2cm}
\epsfxsize10cm
\epsffile{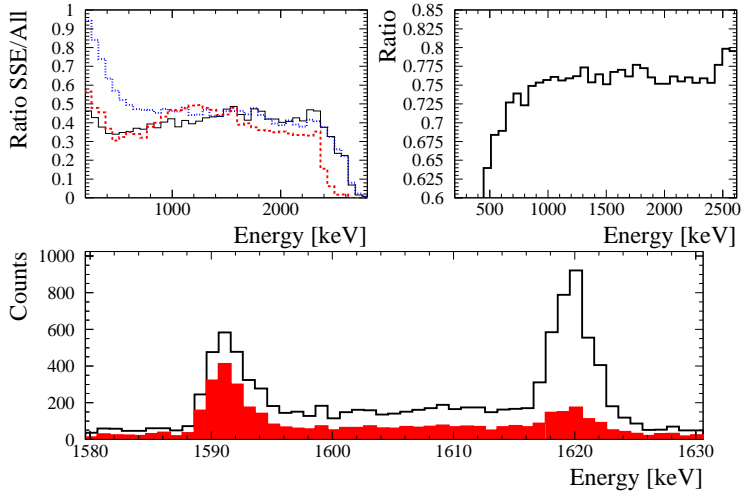}
\caption{\label{ratio} Upper Panel left: Fraction of SSE's in the spectrum
  obtained using Neural Network (solid line), applying the one-parameter
  cut (densely dashed line) and the expectation from the simulation
  (losely dashed  
  line). Upper panel right: Fraction of pulses identified identically by
  Neural Network and one-parameter cut. Lower panel: Measured spectrum 
  in the energy region of the reference pulses (data independent of
  the training sample). The filled histogram corresponds to the
  identified SSE pulses. The open histogram shows the spectrum of all
  events.}
\end{figure}

\section{Network results}
We recorded $\sim$ 20.000 events of each kind (1592~keV Double-Escape
line and 1621 keV total absorption peak) with every detector of
the \HdMo~in the Gran Sasso Underground laboratory. After arranging
these pulses in a library, they were used
to train the networks. Since the Pulseshapes are dependent on detector
parameters like size and form of the crystal, an own network had to be 
configured for every detector used in the PSA.

The Neural Networks used in this analysis consisted out of three layers:
An input layer out of 180 neurons, a hidden layer with 90 neurons and
an output layer with a single neuron.

To check the network also during the training process, we monitored
the evolution of the network with time, i.e. with the number of
presented pulses. This is shown in Fig. \ref{net_evolution} for one of 
the
enriched detectors: In the upper panel the output of the Network 
for the training peaks of the MSE and SSE lines are depicted
separately as a function of time. It is evident that the network
stabilizes after $\sim$ 400.000 
presented pulses (i.e. each pulse has been passed to 
the network $\sim$ 10 times) and is able to
distinguish between the two types of events. The fact that the network 
is able to identify the contamination of wrong pulses (from the
admixture of SSE's to the MSE sample and of MSE's to the SSE sample) in
the separate
libraries gives us further confidence in the power of the method. This
is shown in the lower panel. Here the fraction of identified SSE Pulses
within the two samples of the training pulses is drawn as a function of
presented pulses. The shaded areas give the one sigma region for the
expectation obtained from simulations (see section 5). Note that the
fraction of 'wrong pulses' in the libraries is not an input parameter
to the training process. This behaviour is obtained solely by the
presentation of the reference pulses, i.e. the Neural Network itself
recognizes the contamination without previous knowledge.

From Fig. \ref{net_evolution} it is obvious that
further training of the
network is meaningless after a certain limit, the obtained separation
into MSE and 
SSE is not stabilizing further after $\sim$ 400.000 presented pulses. 
The separation is fluctuating around its mean value from here on.
This is probably due to the fact that 
a non negligible amount of wrong pulses is contained in the training
samples. In 
principle it would be possible to remove a large amount of this
contamination since the network identifies wrong pulses within the
samples itself. However it seemed too dangerous to use this method
for further training since this could give rise to sytematic effects.

\begin{figure}[h]
\hspace*{2.8cm}
\epsfxsize10cm
\epsffile{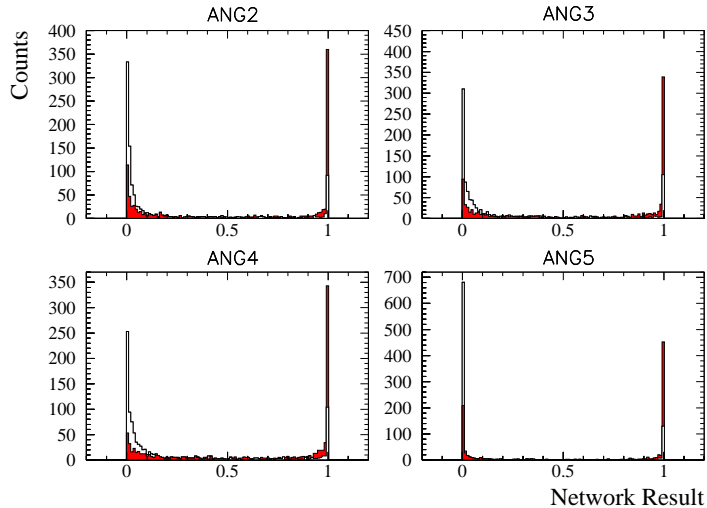}
\caption{\label{seperation} Result of the trained networks tested on
  independent data, i.e. on pulses which were not used for the
  training process. The blank histograms correspond to results of events 
  from the SSE-library (expected result: 0), the filled histograms to
  events from the MSE-library (expected result: 1). Note that the
  network identifies the contamination with wrong pulses in the
  refernce libraries correctly.}
\end{figure}

Once the training process is finished, it is important to check the
obtained results with independent data not used in the training
process. We saved one thousand events of each kind for every detector
to do this test. The result is seen in
Fig. \ref{seperation}. As is evident, also for the independent data the 
separation works very well (for a quantitative analysis see section 7).

In Fig. \ref{seperation} it is visible that for a non-negligible  
fraction of the pulses an output $y_{o} $ between 0.1 and 0.9 is
returned from the network, i.e. the pulses are not properly attributed 
to a definite type. 
The fraction of these pulses is $\sim$ 20\% for all the detectors.
This quantity can be identified as the efficiency of the
separation. However, since we
have further information from the simulation, we use this fact first to
adjust the outcome of the networks to the expectations from the simulation.
We define a cut value $\zeta$ so that all pulses with $y_{o} < \zeta$
are identified as SSE. 
To adjust the network to the expected result we vary $\zeta$  
and perform a least square fit for the simulated
and measured SSE ratios over the whole energy range above 500~keV.
Having found the best fit, this cut is applied to the network result 
and thus the SSE-spectra are obtained. 
Note that separation between the two type of events has been
obtained this way in Fig. \ref{net_evolution}.
The result with this cut value $\zeta$ is then used to calculate the
efficiencies $e_{s}$ and $e_{m}$ of the correct identification of SSE
pulses and MSE pulses (see section 7).

In the lower panel of Fig. \ref{ratio} the obtained result for the
energy-region around the reference pulses is shown. Most events in the
Double-Escape-line are recognized as SSE's. Only a small fraction from 
the background contributes to a contamination of MSE's. Also the
1621~keV peak is recognized correctly to consist mainly of MSE's.

\section{Comparison with the simulation and the old parameter cut}

To compare the results of simulation and measurement directly,
the measured and expected ratios of SSE's in the spectrum 
as a function of energy are
shown in Fig. \ref{ratio} together with the result from the old
one-parameter method. It is evident that the result from the neural
netwok is satisfactory over the whole energy range above $\sim$ 500 keV. 
Only below $\sim$1000~keV there is a noticable
difference between the neural network method and the old method. Here 
the old cut yields too many SSE pulses.
Note that especially in the energy region interesting for \On~
(2000~keV-2080~keV) the agreement of the two techniques is very good.

\begin{table}
\begin{tabular}{c|cc|cc}
\hline
\textbf{Detector} & \textbf{Ratio measured} & \textbf{Ratio simulated}
& \textbf{Ratio measured} & \textbf{Ratio simulated}\\
&\multicolumn{2}{c}{\textbf{Double-Escape Peak 1592~keV}} \vline& \multicolumn{2}{c}{\textbf{1621 keV Peak}}\\
\hline
ANG2 & 70.9$\pm$2.7 & 71.7$\pm$7.7 & 28.3$\pm$1.7 & 18.0$\pm$4.8 \\
ANG3 & 72.4$\pm$2.7 & 75.1$\pm$7.8 & 29.2$\pm$1.7 & 17.5$\pm$4.7 \\
ANG4 & 72.2$\pm$2.7 & 74.8$\pm$7.3 & 29.9$\pm$1.7 & 18.5$\pm$3.4 \\
ANG5 & 76.0$\pm$2.8 & 76.4$\pm$8.5 & 28.7$\pm$1.7 & 17.4$\pm$4.5 \\
\hline
\end{tabular}\caption{\label{lines} Fraction of identified SSE events
  in the peak
  areas of the Double-Escape 1592~keV line and the total absorption
  1621~keV peak and their 
  expectations from the simulation.}
\end{table}

In Tab. \ref{lines} the fraction of identified SSE's in the
Double-Escape peak   
is listed for the four detectors together with the expected results
from the simulation. As evident, the measured results are in good agreement
with the expectation for the Double-Escape peak.
The situation for the measured SSE fraction in the 1621 keV peak is
slightly different. Since the efficiency $e_{m}$ for correct
identification of MSE's is not 100\%, the actual measured SSE fraction 
within this peak is somewhat higher than the expected fraction from
the simulation.
Once the real fraction of SSE's in a certain energy region is known
through e.g. a simulation, it is easy to calculate the efficiencies
of the recognition:
\begin{equation}
e_{s}=\frac{1}{\gamma_{s}}+\left(1-\left(\frac{a}{s}\right)_{SSE}\right)\frac{\frac{1}{\gamma_{s}}-\frac{1}{\gamma_{m}}}{\left(\frac{a}{s}\right)_{SSE}-\left(\frac{a}{s}\right)_{MSE}}
\end{equation}
and 
\begin{equation}
e_{m}=1-\frac{\frac{1}{\gamma_{s}}-\frac{1}{\gamma_{m}}}{\left(\frac{a}{s}\right)_{SSE}-\left(\frac{a}{s}\right)_{MSE}}
\end{equation}

where $\gamma_{m}=s_{MSE}$/S$_{MSE}$ is the ratio of real SSE events in the
1621~keV peak to events identified as SSE's by the network within the
peak, $\gamma_{s}=s_{SSE}$/S$_{SSE}$ is the according fraction for the
Double-Escape-peak and
$(\frac{a}{s})_{MSE}$ and $(\frac{a}{s})_{SSE}$ are the
simulated ratios of all events to SSE events in the given energy
region. The total efficiency $e_{tot}$ of the method is then given by
the square root of the product of the two single efficiencies.

The obtained efficiencies for the four networks in the \HdMo~
are listed in Tab. \ref{efficiencies}. 
\begin{table}
\hspace*{3cm}
\begin{tabular}{cccc}
\hline
\textbf{Detector}&\textbf{$e_{s}$} & \textbf{$e_{m}$} & \textbf{$e_{tot}$}\\
\hline
ANG2 & 0.93$\pm$0.27 & 0.86$\pm$0.25 & 0.90$\pm$0.38\\
ANG3 & 0.91$\pm$0.26 & 0.84$\pm$0.24 & 0.87$\pm$0.37\\
ANG4 & 0.91$\pm$0.19 & 0.84$\pm$0.17 & 0.87$\pm$0.27\\
ANG5 & 0.95$\pm$0.27 & 0.85$\pm$0.24 & 0.90$\pm$0.38\\
\hline
\end{tabular}
\caption{\label{efficiencies}Efficiencies for correct SSE and MSE
  identification for the detectors of the \HdMo~by neural network.}
\end{table}

Obviously an efficient separation of MSE and SSE pulses can be
accomplished with Neural Networks. In principle it is possible to
correct the result of the network through the known efficiencies
with the relation
\begin{equation}
s=\frac{S-(1-e_{m})A}{(e_{s}+e_{m}-1)}.
\end{equation}
Here S is the number of SSE's identified by the network, A is the
total number of events and $s$ is the real amount of SSE in the sample 
A. In our case the correction yields a
smaller SSE rate $s$ then actually obtained with the neural networks S,
since the number of MSE's identified as SSE's is larger than the 
number of SSE's identified as MSE's. 
\begin{equation}
(1-e_{m})(A-s) > (1-e_{s})s.
\end{equation}

However, we decided not to make use of this correction.
Since the obtained efficiencies are high, the
correction would be only of the order of 30\% in the case of a large
ratio $\frac{a}{s}$, as realized for total absorption peaks. The
correction would be of the order of $\sim$10\% for the expected ratio
in the \On~energy range.
The fact that we do not apply the correction makes the use of the new
technique a conservative method.

From the ratio of identified SSE
pulses in the energy region between 2000~keV and 2080~keV it is
expected that the background in the calibration spectrum 
can be further
reduced by a factor of 2.67$\pm$0.05 which is in good
agreement with the results obtained from the simulation which yields
a reduction factor of 2.78$\pm$0.01 and the one parameter-cut, which 
gives a reduction by a factor of 2.53$\pm$0.05 The slightly smaller
value in the measurement is the effect of the efficiencies
for the recognition.
We expect a similar reduction for the \HdMo .


To finally check the compatibility of the two methods, in the right diagram of 
Fig. \ref{ratio}
we show the fraction of pulses from the library which were attributed
the same type from both methods as a function of energy.
With the efficiencies given above and the efficiencies of the old
method we expect a fraction of $\sim$ 70 \% to be identified equaly. 
Indeed $\sim$75\% of the events are classified equally.

\section{Conclusion}
We developed a new method to distinguish between multiple scattered
and single interaction events in HPGe-detectors on the basis of Neural
Networks. We 
showed that this technique is capable of distinguishing between the
two types of events with a very high efficiency. The
comparison with a simulation performed for this purpose confirms these 
results.

\section{Aknowledgement}
B. M. would like to thank Ch. Gund for help with the
extension of the GEANT3.21 code. B.~M. is supported by the
Graduiertenkolleg of the University of Heidelberg.

\end{document}